\documentclass[fleqn,twoside]{article}
\usepackage{espcrc2}

\usepackage{graphicx}
\usepackage[figuresright]{rotating}
\usepackage{epsfig}
\usepackage{graphicx}

\newcommand{\AmS}{{\protect\the\textfont2
  A\kern-.1667em\lower.5ex\hbox{M}\kern-.125emS}}

\def\beq{\begin{equation}}
\def\eeq{\end{equation}}
\def\bea{\begin{eqnarray}}
\def\eea{\end{eqnarray}}
\def\barr{\begin{array}}
\def\earr{\end{array}}

\def\op{{\mathcal{O}}}

\def\exponential{{\mathrm{e}}}

\title{Heavy meson chiral perturbation theory in finite volume\thanks{
INT-PUB 04-22, NT@UW-04-22, UW/PT 04-18, work supported by
DoE grants DE-FG03-96ER40956, DE-FG03-97ER41014 and DE-FG03-00ER41132.}}

\author{C.-J. David Lin\address{Department of Physics,
University of Washington, Seattle, WA 98195-1560, U.S.A.}\address
{Institute for Nuclear Theory,
University of Washington, Seattle, WA 98195-1550, U.S.A.}}
       
\begin{document}

\begin{abstract}
We present the first step towards the estimation of 
finite volume effects in heavy-light meson systems
using heavy meson chiral perturbation theory.  We 
demonstrate that these effects can be amplified in both
light-quark and heavy-quark mass extrapolations (interpolations)
in lattice calculations.  As an explicit example, we perform a
one-loop calculation for the neutral $B$ meson mixing system and
show that finite volume effects, which can be comparable with
currently quoted errors, are not negligible in both quenched
and partially quenched QCD.
\vspace{1pc}
\end{abstract}

\maketitle

\section{INTRODUCTION}
\label{sec:intro}
Advances in lattice QCD have lead to the prospect of numerical calculations for
hadronic matrix elements with high precision, which will have
significant impact on the progress of particle physics in the near future.  One
of the areas where lattice QCD plays an essential r\^{o}le is the search for 
physics beyond the Standard Model via
over-constraining the $b{-}d$ unitarity triangle in the 
Cabibbo-Kobayashi-Maskawa (CKM) matrix.
This requires the control of systematic
errors at the percentage level for certain hadronic matrix elements in 
heavy-light meson mixing
and decays.

Here we present the first step towards the estimation of
finite-volume effects in heavy-light meson systems 
\cite{Arndt:2004bg} in the framework of heavy
meson chiral perturbation theory (HM$\chi$PT)
\cite{Burdman:1992gh} \cite{Wise:1992hn} \cite{Yan:1992gz} 
\cite{Sharpe:1995qp} \cite{Booth:1994hx}, with first 
order $1/M_{P}$ and chiral corrections, assuming
the mass hierarchy
\beq
\label{eq:mass_hierarchy}
 M_{\mathrm{GP}} \ll \Lambda_{\chi} \ll M_{P} ,
\eeq
where $M_{\mathrm{GP}}$ is the mass of any Goldstone particle, 
$M_{P}$ is the mass of the heavy-light meson, and 
$\Lambda_{\chi}$ is the chiral symmetry breaking scale.  Under this
assumption, we discard corrections of the size
$M_{\mathrm{GP}}/M_{P}$.
%%
%\beq
%\label{eq:mgp_over_mp}
% \frac{M_{\mathrm{GP}}}{M_{P}} .
%%\eeq
% 
Concerning the finite volume, we work with the
condition that
\beq\label{eq:MpiL_big}
 M_{\mathrm{GP}} L \gg 1,
\eeq
where $L$ is the spatial extent of the cubic box.  The temporal
direction of the lattice is taken to be infinite.

The main task of this work is to study the volume effects due to the presence
of the scales 
\beq
\label{eq:Delta_def}
 \Delta = M_{P^{\ast}} - M_{P} ,
\eeq
and
\beq
\label{eq:delta_s_def}
 \delta_{s} = M_{P_{s}} - M_{P} ,
\eeq
where $P^{\ast}$ and $P$ are the heavy-light vector and pseudoscalar mesons 
containing a $u$ or $d$ anti-quark\footnote{We work in the isospin limit in
this paper.}, and $P_{s}$ is the 
heavy-light pseudoscalar meson with an $s$ anti-quark. The
scale $\Delta$ appears due to the breaking of heavy quark spin symmetry
that is of $\op(1/M_{P})$ and $\delta_{s}$ comes from light flavour $SU(3)$ 
breaking in the heavy-light meson masses\footnote{Under the assumption
of Eq.~(\ref{eq:mass_hierarchy}), $\Delta$ is independent of the
light quark mass, and $\delta_{s}$ does not contain any $1/M_{P}$
corrections, to the order
we are working.}.  These scales are comparable to the pion mass
in the real world and in lattice simulations.  Therefore it is important
to understand how they combine with the infra-red scales $1/L$ and 
$M_{{\mathrm{GP}}}$ in finite volume.

\section{HEAVY-LIGHT MESONS IN FINITE VOLUME}
\label{sec:FV}
Finite volume effects in heavy-light meson systems are dominated
by the Goldstone particles which couple to the heavy meson and
can ``wrap around the world''. In HM$\chi$PT, 
the heavy-light pseudoscalar $P$ meson can scatter into 
the vector $P^{\ast}$ meson by emitting a pion.  
In the limit the heavy quark
mass is infinite, where the heavy-quark spin symmetry is exact,
both $P$ and $P^{\ast}$ mesons are on-shell static sources
and there is a velocity superselection rule.  This can be
seen from their propagators
\beq
 \frac{i}{2 (v\cdot k + i\epsilon)} \mbox{ },\mbox{ }\mbox{ }
 \frac{-i (g_{\mu\nu} - v_{\mu}v_{\nu})}{2 (v\cdot k + i\epsilon)} ,
\eeq
which are proportional to $\theta(t)\delta^{(3)}(\vec{x})$ in
position space.  In this situation, finite
volume effects can only be of the form 
exp($-|\vec{n}|M_{{\mathrm{GP}}}L$)/($|\vec{n}|M_{{\mathrm{GP}}}L$), 
which is the position-space.propagator for a Goldstone
particle wrapping around the world $n_{i}$ times in the spatial
direction $i$.

Away from the strict heavy-quark limit, there is a mass
difference, $\Delta\sim 1/M_{P}$, between $P^{\ast}$ and $P$ mesons.
This mass shift, in the presence of the velocity superselection
rule, renders the 
propagator of the $P^{\ast}$ meson into the form
\beq
 \frac{-i (g_{\mu\nu} - v_{\mu}v_{\nu})}{2 (v\cdot k - 
  \Delta + i\epsilon)} ,
\eeq
and brings it off-shell with the virtuality $\Delta$.
The time uncertainty conjugate to this virtuality, 
$\delta t\sim 1/\Delta$, restricts the period
during which the Goldstone particles can propagate to wrap
around the world, hence alters volume
effects. Therefore in the class of diagrams involving the 
$P{-}P^{\ast}{-}\pi$ coupling in HM$\chi$PT, 
volume effects decrease
with increasing $\Delta$ and can only 
depend on $M_{{\mathrm{GP}}}/\Delta$.

\begin{figure}[htb]
\begin{center}
\includegraphics[width=6.5cm]{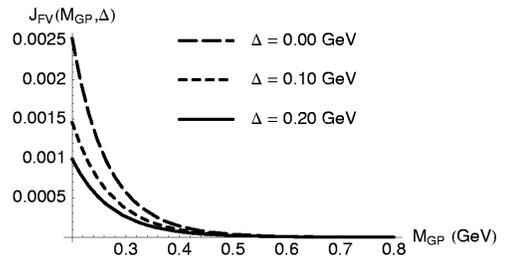}
\vspace{-1cm}
\caption{\label{fig:JFV}{\sl $J_{\mathrm{FV}}(M_{\mathrm{GP}},
\Delta)$ at $L = 2.5$ fm,
plotted as 
a function of $M_{\mathrm{GP}}$. 
The Goldstone mass $M_{\mathrm{GP}} = 0.197$ GeV corresponds
to $M_{\mathrm{GP}} L = 2.5$, and $M_{\mathrm{GP}} = 0.32$ GeV corresponds
to $M_{\mathrm{GP}} L = 4$.}}
\end{center}
\end{figure}

The above physical picture can be observed in a 
momentum-space calculation by considering a typical
sum in one-loop HM$\chi$PT,
\bea
 & &{\mathcal{J}}(M_{\mathrm{GP}},\Delta) 
 =-i  \frac{1}{L^{3}}
 \sum_{\vec{k}} \int \frac{d k_{0}}{2\pi}\nonumber\\
\label{eq:calJ}
 & &\mbox{ }\mbox{ }\mbox{ }\mbox{ }\mbox{ }
  \times\frac{1}{(k^{2}-M^{2}_{\mathrm{GP}} + i\epsilon )
 (v\cdot k - \Delta + i\epsilon )}  ,
\eea
where the spatial momentum $\vec{k}$ is quantised in finite 
volume as $2\pi \vec{i}/L$,
with $\vec{i}$ being a three dimensional integer vector.
In the infinite-volume limit, the sum ${\mathcal{J}}$
becomes an integral with $\sum_{\vec{i}}/L^{3}$ 
replaced by $\int d^{3}k/(2\pi)^{3}$.
Using the Poisson summation formula, it can be shown
that finite volume effects in this sum are
(with $n=|\vec{n}|$)
\beq
\label{eq:JFV_asymp}
 J_{\mathrm{FV}}(M_{\mathrm{GP}},\Delta) = 
\sum_{\vec{n}\not=\vec{0}} \left ( \frac{1}{8\pi n L}\right )
{\mathrm{e}}^{- n M_{\mathrm{GP}} L} 
{\mathcal{A}} ,
\eeq
in the 
asymptotic limit $M_{{\mathrm{GP}}} L \gg 1$, where
\beq
\label{eq:JFV_asymptotic}
{\mathcal{A}} =\exponential^{(z^{2})} \big [ 
1 - {\mathrm{Erf}}(z)\mbox{ }\big ]
%&+&\left (\frac{1}{n M_{\mathrm{GP}} L} \right ) \bigg [
% \frac{1}{\sqrt{\pi}} \left ( \frac{z}{4} - 
%\frac{z^{3}}{2}\right )
%\nonumber\\
% & &+\frac{z^{4}}{2}\exponential^{(z^{2})} 
% \big [ 1 - {\mathrm{Erf}}(z)\mbox{ }\big ]
%\bigg ]\nonumber\\
%\label{eq:JFV_asymptotic}
%%& &
%%-\left (\frac{1}{n M_{\mathrm{GP}} L} \right )^{2}\bigg [
%%\frac{1}{\sqrt{\pi}}\left ( \frac{9z}{64} - 
%%\frac{5z^{3}}{32}
%%  +\frac{7z^{5}}{16} + \frac{z^{7}}{8} \right )
%%-\left ( \frac{z^{6}}{2} + \frac{z^{8}}{8}\right )
%%\exponential^{(z^{2})} \big [ 1 - {\mathrm{Erf}}(z)\mbox{ }\big ]
%%\bigg ]
%%%\nonumber\\
%%%& &
+\op\left (\left [ \frac{1}{n M_{\mathrm{GP}} L}\right ]\right ) ,
\eeq
%
%\end{widetext}
with
\beq
z \equiv \left (\frac{\Delta}{M_{\mathrm{GP}}}\right ) 
\sqrt{\frac{n M_{\mathrm{GP}}L}{2}} .
\eeq
The quantity ${\mathcal{A}}$ is the alteration of finite volume
effects due to the presence of a non-zero $\Delta$.  
It multiplies the factor
exp($-n M_{\mathrm{GP}} L$), which results from 
the Goldstone particles wrapping around the world.  Notice
that it is possible to analytically compute the higher
order corrections of ${\mathcal{A}}$ in powers of 
$1/(n M_{{\mathrm{GP}}} L)$ to achieve any desired
numerical precision.  The function $J_{{\mathrm{FV}}}$
is plotted in Fig. \ref{fig:JFV} at $L = 2.5$ fm.  
\vspace{-0.5cm}
\vspace{-0.5cm}
This plot shows that $J_{{\mathrm{FV}}}$ is strongly dependent
on $\Delta$, hence $M_{P}$, and it decreases as $\Delta$
increases.

\section{CALCULATION FOR $B^{0}{-}\bar{B}^{0}$ MIXING}
\label{sec:BBbar}
\begin{figure}[htb]
\begin{center}
\includegraphics[width=7.5cm]{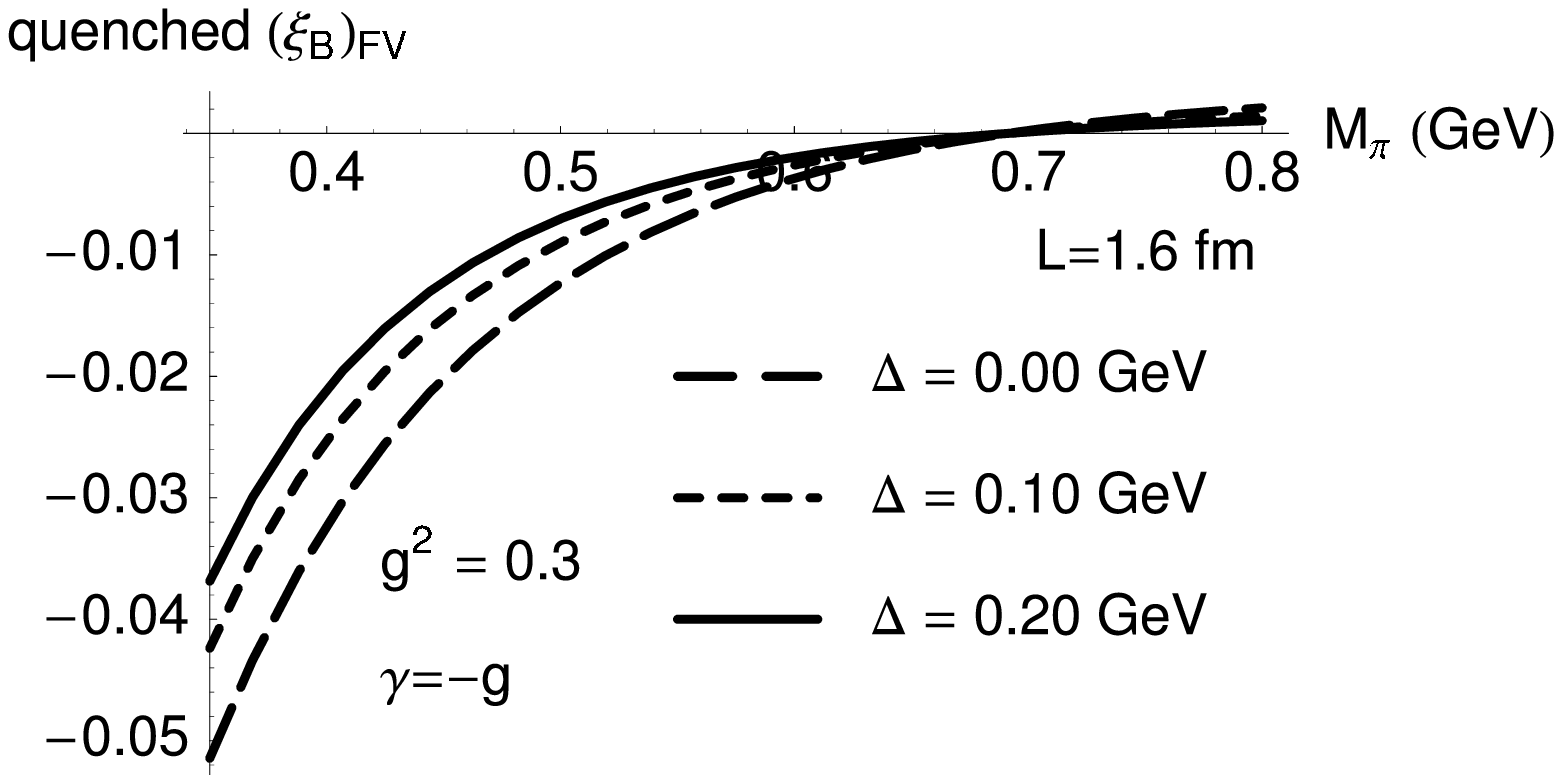}
\includegraphics[width=7.5cm]{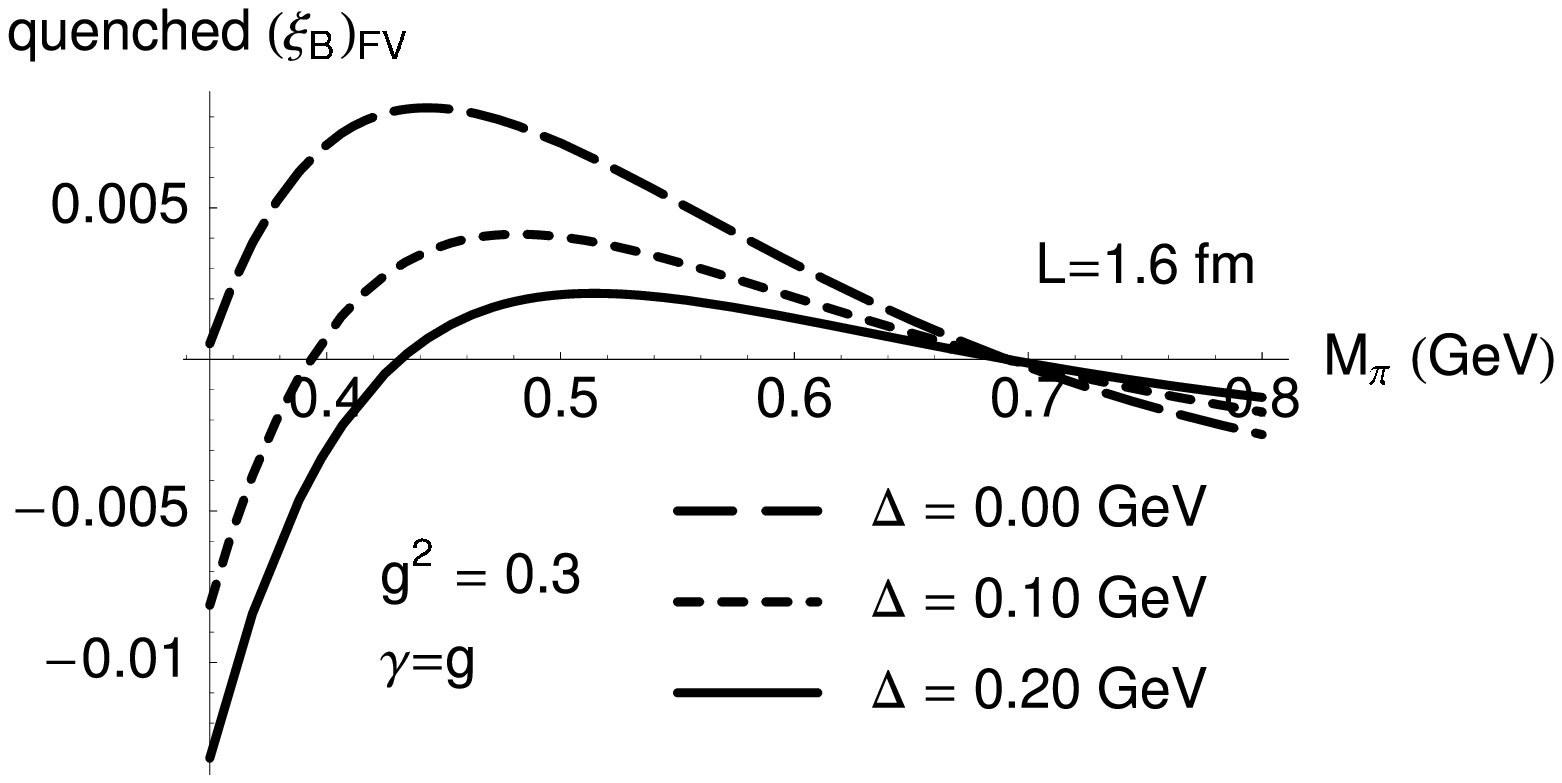}
\includegraphics[width=7.5cm]{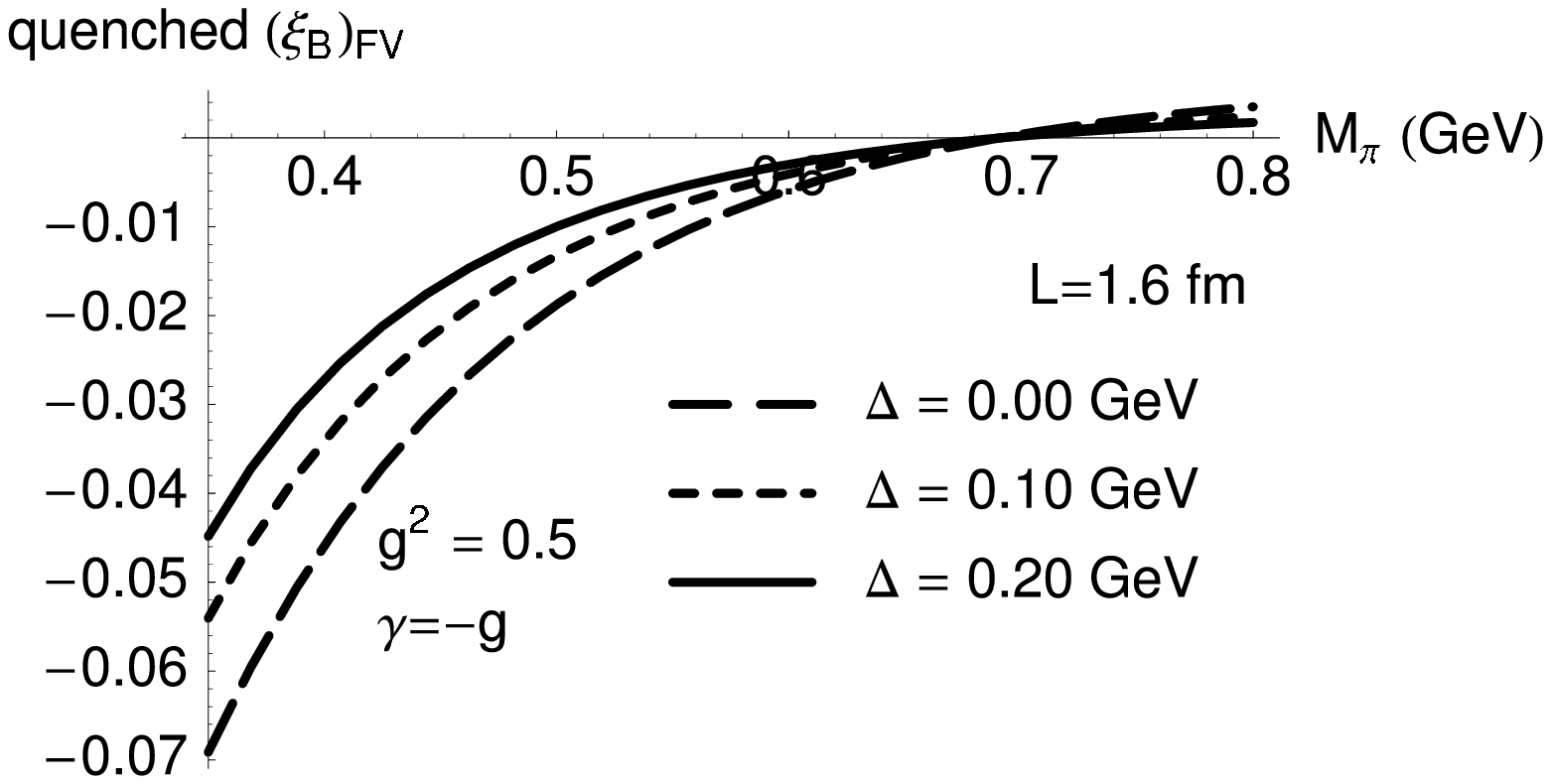}
\includegraphics[width=7.5cm]{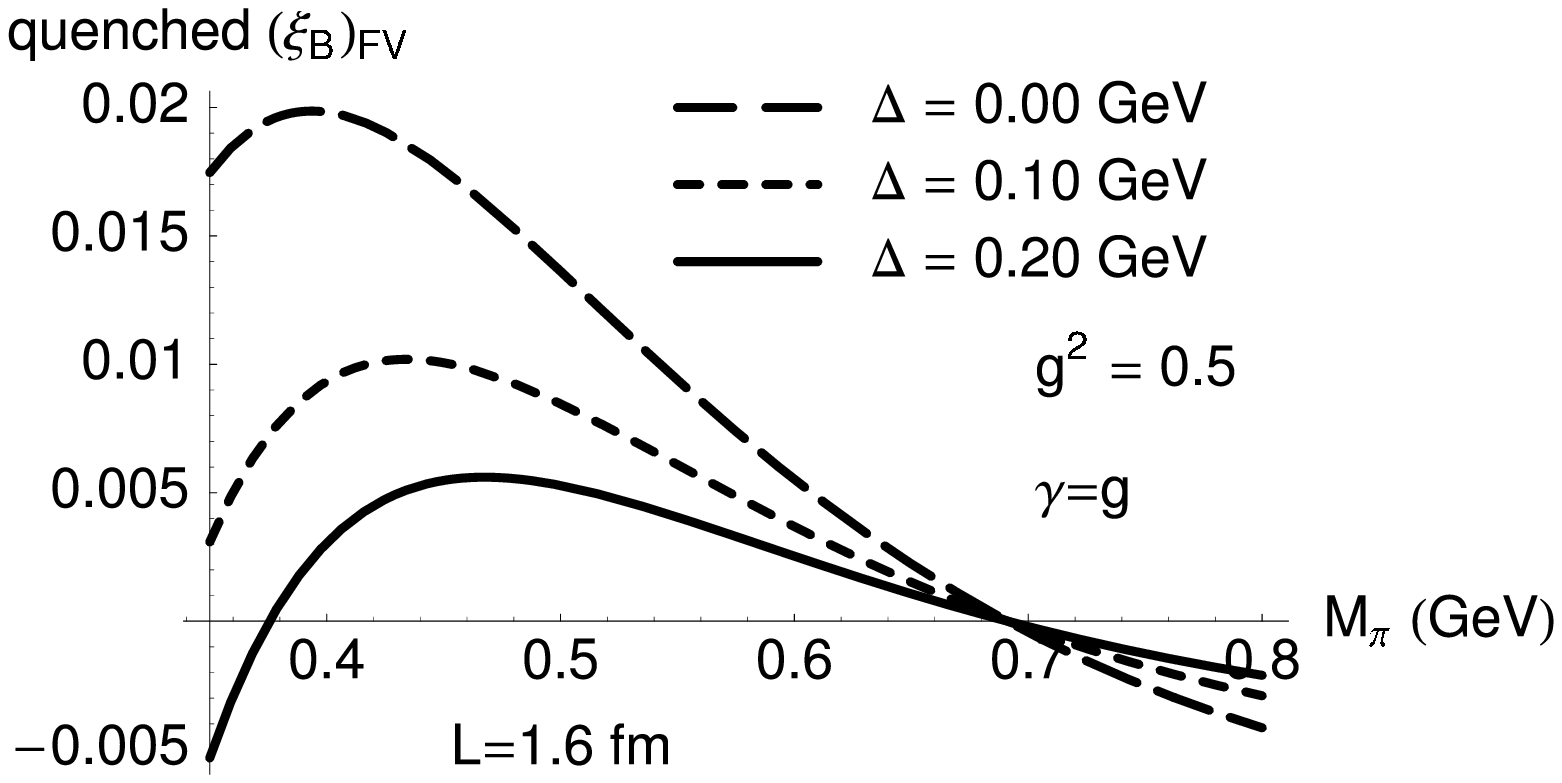}
\vspace{-1cm}
\caption{\label{fig:Q_BBs_over_BB}{\sl $(\xi_{B})_{\mathrm{FV}}$ 
in QQCD plotted against
$M_{\pi}$, with $L=1.6$ fm, strange-quark mass set to its
physical value and choices of the couplings 
$g$ (the $P{-}P^{\ast}{-}\pi$ coupling) and 
$\gamma$ (the $P{-}P^{\ast}{-}\eta^{\prime}$ coupling). 
The pion mass $M_{\pi} = 0.35$ GeV corresponds
to $M_{\pi} L = 2.8$, and $M_{\pi} = 0.5$ GeV corresponds
to $M_{\pi} L = 4$ in this plot.  
We set $\alpha=0$ and $M_{0}=700$ MeV.  Notice that at
$\Delta\sim$ 50 MeV and 150 MeV at physical $M_{B}$
and $M_{D}$ respectively.}}
\end{center}
\end{figure}

We have performed a one-loop calculation for the 
$B^{0}_{(s)}{-}\bar{B}^{0}_{(s)}$ mixing system, including the 
$B_{B_{(s)}}$ parameters and decay constants\footnote{Notice that
the results presented here can be used to analyse the box-diagram
contribution to the matrix elements of $D^{0}{-}\bar{D}^{0}$ mixing
as well.} in full, quenched and $N_{f}=2+1$ 
partially-quenched QCD.  Especially, we investigate volume effects in 
the SU(3) breaking ratios:
\beq
\label{eq:xi_f_B}
 \xi_{f} = \frac{f_{B_{s}}}{f_{B}}\mbox{ }{\mathrm{and}}\mbox{ }\mbox{ }
 \xi_{B} = \frac{B_{B_{s}}}{B_{B}} ,
\eeq
which are important inputs for the global CKM fit.
Furthermore, we define 
\beq
 (\xi_{f})_{\mathrm{FV}}\mbox{ }{\mathrm{and}}\mbox{ }
 (\xi_{B})_{\mathrm{FV}}
\eeq
to be the contributions from volume effects.  They are estimated
by calculating
the volume-dependent one-loop 
corrections {\it with respect to the lowest-order values} of 
$f_{B_{s}}$ ($B_{B_{s}}$) and $f_{B}$ ($B_{B}$), then taking the difference
between the results.  We find that volume effects are more
salient in $\xi_{B}$ than $\xi_{f}$.  For existing and future lattice
calculations,  $(\xi_{B})_{\mathrm{FV}}$ are typically $\sim 5\%$ for
quenched,  $\sim 4\%$ for partially quenched and $\sim 2\%$ for full
QCD, {\it in the parameter space where lattice simulations are carried
out, and they can be amplified in both light-quark and heavy-quark
mass extrapolations (interpolations)}.  An example in quenched
QCD is shown in 
Fig. \ref{fig:Q_BBs_over_BB}.  From these plots, it is clear
that finite volume effects have strong dependence on both
light-quark and heavy-quark masses and can exceed the currently
quoted errors on $\xi_{B}$.

\vspace{-0.5cm}
\vspace{1cm}
%
%

%%%%%%


\begin{thebibliography}{9}

%\cite{Arndt:2004bg}
\bibitem{Arndt:2004bg}
D.~Arndt and C.-J.~D.~Lin,
%``Heavy meson chiral perturbation theory in finite volume,''
Phys.\ Rev.\ D {\bf 70}, 014503 (2004)
[arXiv:hep-lat/0403012].
%%CITATION = HEP-LAT 0403012;%%

%\cite{Burdman:1992gh}
\bibitem{Burdman:1992gh}
G.~Burdman and J.~F.~Donoghue,
%``Union of chiral and heavy quark symmetries,''
Phys.\ Lett.\ B {\bf 280}, 287 (1992).
%%CITATION = PHLTA,B280,287;%%

%\cite{Wise:1992hn}
\bibitem{Wise:1992hn}
M.~B.~Wise,
%``Chiral perturbation theory for hadrons containing a heavy quark,''
Phys.\ Rev.\ D {\bf 45}, 2188 (1992).
%%CITATION = PHRVA,D45,2188;%%

%\cite{Yan:1992gz}
\bibitem{Yan:1992gz}
T.~M.~Yan {\it et al.},
%``Heavy quark symmetry and chiral dynamics,''
Phys.\ Rev.\ D {\bf 46}, 1148 (1992)
[Erratum-ibid.\ D {\bf 55}, 5851 (1997)].
%%CITATION = PHRVA,D46,1148;%%

%\cite{Sharpe:1995qp}
\bibitem{Sharpe:1995qp}
S.~R.~Sharpe and Y.~Zhang,
%``Quenched Chiral Perturbation Theory for Heavy-light Mesons,''
Phys.\ Rev.\ D {\bf 53}, 5125 (1996)
[arXiv:hep-lat/9510037].
%%CITATION = HEP-LAT 9510037;%%

%\cite{Booth:1994hx}
\bibitem{Booth:1994hx}
M.~J.~Booth,
%``Quenched chiral perturbation theory for heavy - light mesons,''
Phys.\ Rev.\ D {\bf 51}, 2338 (1995)
[arXiv:hep-ph/9411433].
%%CITATION = HEP-PH 9411433;%%

\end{thebibliography}
\end{document}